\documentclass[12pt]{article}
\usepackage[dvips]{graphicx}
\usepackage{amsmath,amssymb,amsthm,bm}
\bmdefine{\BX}{{\bm X}}
\bmdefine{\Bx}{{\bm x}}
\bmdefine{\By}{{\bm y}}
\bmdefine{\BY}{{\bm Y}}
\bmdefine{\Bz}{{\bm z}}
\bmdefine{\Ba}{{\bm a}}
\bmdefine{\Bd}{{\bm d}}
\bmdefine{\Bu}{{\bm u}}
\bmdefine{\Bm}{{\bm m}}
\bmdefine{\Bi}{{\bm i}}
\bmdefine{\Bc}{{\bm c}}
\bmdefine{\BT}{{\bm T}}
\bmdefine{\Bp}{{\bm p}}
\bmdefine{\Bs}{{\bm s}}
\bmdefine{\Btheta}{{\bm \theta}}
\bmdefine{\Bmu}{{\bm \mu}}
\bmdefine{\Bt}{{\bm t}}
\bmdefine{\Blambda}{{\bm \lambda}}
\bmdefine{\Bbeta}{{\bm \beta}}
\bmdefine{\Bpsi}{{\bm \psi}}
\bmdefine{\Bpi}{{\bm \pi}}
\bmdefine{\Bv}{{\bm v}}
\bmdefine{\Bzero}{{\bm 0}}
\bmdefine{\Bone}{{\bm 1}}
\bmdefine{\Bvarepsilon}{{\bm \varepsilon}}

\bmdefine{\BI}{{\rm {\bm I}}}
\bmdefine{\BV}{{\rm {\bm V}}}

\newtheorem{theorem}{Theorem}[section]

\theoremstyle{definition}

\newtheorem{definition}[theorem]{Definition}

\title{
Use of primary decomposition of polynomial ideals 
arising from indicator functions 
to enumerate orthogonal fractions
}
\author{Satoshi Aoki%
\thanks{Graduate School of Science, Faculty of Science, Kobe
University.}
and 
Masayuki Noro%
\thanks{Department of Mathematics, College of Science, Rikkyo University}
}
\date{}

\begin{document}
\maketitle

\begin{abstract}
\noindent
A polynomial indicator function of designs is first introduced by
 Fontana {\it et al}. (2000) for two-level cases.
They give the structure of the indicator functions, especially the 
 relation to the orthogonality of designs. These results are 
 generalized by Aoki (2019) for general multi-level cases. 
As an application of these results, we can enumerate all orthogonal
 fractional factorial designs with given size and orthogonality using
 computational algebraic software. For example, Aoki (2019) gives 
 classifications of orthogonal fractions of $2^4\times 3$ designs with
 strength $3$, which is derived by simple eliminations of
 variables. However, the
 computational feasibility of this naive approach depends on the size
 of the problems. In fact, it is reported that the computation of 
orthogonal fractions of $2^4\times 3$ designs with
 strength $2$ fails to carry out in Aoki (2019). In this paper, using 
 the theory of primary decomposition, we enumerate and classify 
orthogonal fractions of $2^4\times 3$ designs with
 strength $2$. We show there are 
$35200$ orthogonal half fractions of $2^4\times 3$ designs with strength
$2$, classified into $63$ equivalent classes. 

\end{abstract}

\section{Introduction}
\label{sec:intro}
An application of Gr\"obner basis theory to problems in the field of 
statistics first 
arises in the design of experiments. In the first paper of this research
field named computational algebraic statistics, Pistone and Wynn reveal the
relation between the identifiability problem of polynomial models on
designs and the ideal membership problem (\cite{Pistone-Wynn-1996}). 
After this breakthrough paper, various applications of algebraic
methods to the problems in the design of experiments are given by
researchers both in the field of statistics and algebra. 
The theory of the indicator function is one of the early, fundamental
results in this field.

The indicator function of a fractional factorial design is first
introduced for the designs of two-level factors
(\cite{Fontana-Pistone-Rogantin-2000}). 
Based on the arguments of \cite{Pistone-Wynn-1996}, 
Fontana, Pistone and Rogantin
(\cite{Fontana-Pistone-Rogantin-2000}) show the one-to-one correspondence
between the design and its indicator function. By this correspondence,
various statistical concept such as orthogonality, resolution and
aberration can be translated to the structure of the indicator
functions. As an application of this result, they enumerate and classify all
orthogonal fractions of $2^4$ and $2^5$ designs 
by solving the system of algebraic equations 
for the coefficients of the
indicator functions. 

This argument is generalized to the designs of arbitrary multi-level
factors in Aoki
(\cite{Aoki-2019}). In \cite{Aoki-2019}, algebraic equations 
for the coefficients of the indicator functions of designs with given
orthogonality are given in general settings. As an application, 
\cite{Aoki-2019} gives the classification of all 
orthogonal fractions of
$2^3\times 3$ designs with strength $2$ and 
orthogonal fractions of
$2^4\times 3$ designs with strength $3$. 
These results are derived by simple eliminations of variables, which is
one of the fundamental applications of Gr\"obner basis theory. However,
the computational feasibility of this naive approach 
depends on the size of the problems and apparently fails to carry out for
problems of large sizes. 
In fact, it is reported in \cite{Aoki-2019} that 
the Gr\"obner basis calculation of the first elimination ideal for 
the orthogonal fractions of $2^4\times 3$ designs with strength
$2$ is difficult to carry out for standard PC. (It is reported that
the calculation does not finish in $1$ week.) This is the reason why the
results for $2^4\times 3$ designs are given 
for strength $3$, not for strength $2$ in \cite{Aoki-2019}.

In this paper, we report that we have broken through the limit of 
\cite{Aoki-2019} by a technique of computational algebraic geometry. 
Using the theory of primary decomposition, we obtain the enumeration and
classification of half orthogonal fractions of $2^4\times 3$ designs
with strength $2$. The result is summarized as follows: there are
$35200$ orthogonal half fractions of $2^4\times 3$ designs with strength
$2$, classified into $63$ equivalent classes. This result is shown in
detail in
Section \ref{sec:classification}. The contents of the remaining sections
of this paper are as
follows. In Section \ref{sec:indicator-function}, we summarize the
theory of the indicator functions and describe the problems we
consider. In Section 
\ref{sec:primary-decomposition}, we give some definitions and facts on
the primary decomposition, and explain our approach. 
In Section \ref{sec:discussion}, we give some discussion. 


\section{The indicator functions of fractional factorial designs}
\label{sec:indicator-function}
In this section, we give necessary tools and results on the indicator
functions of fractional factorial designs. Mostly, we follow the 
notations and
definitions of \cite{Aoki-2019}. 
The existence, uniqueness
and structure of the indicator functions are derived from the results of 
\cite{Pistone-Wynn-1996}. 

Let $x_1,\ldots,x_n$ be $n$ factors. Let $A_j \subset \mathbb{Q}$ be a
level set of a factor $x_j$ for $j = 1,\ldots,n$, where $\mathbb{Q}$ be
the field of rational numbers. We denote by $r_j = \# A_j$ the
cardinality of $A_j$ and assume $r_j \geq 2$ for $j = 1,\ldots,n$. A
full factorial design of the factors $x_1,\ldots,x_n$ is a direct
product $D = A_1\times
\cdots\times A_n$. 
For later use, we introduce an index set ${\cal I} =
\{(i_1,\ldots,i_n) \in [r_1]\times \cdots \times [r_n]\}$, where we
define $[k] = \{1,2,\ldots,k\}$ for a positive integer $k$, and 
represent $D$ by its design points explicitly as $D = \{\Bd_{\Bi}
\in \mathbb{Q}^n\ :\ \Bi \in {\cal I}\}$. A subset of $D$ is called a
fractional factorial design. A
fractional factorial design $F \subset D$ is written as $F = \{\Bd_{\Bi}
\in D\ :\ \Bi \in {\cal I}'\}$ for a subset ${\cal I}'$ of ${\cal I}$. 

Now we give a definition of an indicator function.
\begin{definition}[\cite{Fontana-Pistone-Rogantin-2000}]
Let $D$ be a full factorial design and 
$F \subset D$ be a fractional factorial design. The indicator
 function of $F$ is a $\mathbb{Q}$-valued function $f$ on $D$ satisfying
\[
 f(\Bd) = \left\{\begin{array}{ll}
1, & \mbox{if}\ \Bd \in F,\\
0, & \mbox{if}\ \Bd \in D\setminus F.
\end{array}
\right.
\]
\end{definition}

To give the structure of the indicator functions, we prepare some algebraic
materials 
of polynomial rings. Let $\mathbb{Q}[x_1,\ldots,x_n]$ be the polynomial
ring with coefficients in $\mathbb{Q}$. For a design $F \subset
\mathbb{Q}^n$, we denote by $I(F)$ the set of polynomials in
$\mathbb{Q}[x_1,\ldots,x_n]$ which are $0$ at every point of $F$, i.e.,
$I(F) = \{f\in\mathbb{Q}[x_1,\ldots,x_n]\ :\ f(\Bd) = 0,\ \forall\ \Bd
\in F\}$. $I(F)$ is an ideal of
$\mathbb{Q}[x_1,\ldots,x_n]$, and is called the design ideal of
$F$. A generator of 
a design ideal of a full factorial design $D$ can be written as 
$G = \left\{x_j^{r_j} - g_j,\ j = 1,\ldots,n\right\}$, where $g_j$ is a
polynomial in $\mathbb{Q}[x_j]$ with the degree less than $r_j,\ j =
1,\ldots,n$, and $I(D)$ is written as $I(D) = \left< G\right>$. This $G$
is a reduced Gr\"obner basis of $I(D)$ for any monomial order. 
We write the set of 
the monomials of $x_1,\ldots,x_n$ that are not divisible by $x_j^{r_j},\
j = 1,\ldots,n$, the leading monomials of $G$, as
\[
 {\rm Est}(D) = \left\{\Bx^{\Ba} = \prod_{j = 1}^nx_j^{a_j}\ :\ \Ba \in
 L\right\},
\]
where
\[
 L = \{\Ba = (a_1,\ldots,a_n) \in \mathbb{Z}_{\geq 0}^n\ :\ 0 \leq a_j
 \leq r_j - 1,\ j = 1,\ldots,n\}
\]
and $\mathbb{Z}_{\geq 0}^n$ is the set of nonnegative integers.

From $D = \{\Bd_{\Bi} \in \mathbb{Q}^n\ :\ \Bi \in {\cal I}\}$ and $L$,
we define a model matrix by $X = \left[\Bd_{\Bi}^{\Ba}\right]_{\Bi \in {\cal
I}; \Ba \in L}$, where $\Bd_{\Bi}^{\Ba} = \prod_{j = 1}^nd_{\Bi
j}^{a_j}$ and $d_{\Bi j}$ is the level of the factor $x_j$ in the
experimental run indexed by $\Bi \in {\cal I}$. By ordering the element
of ${\cal I}$ and $L$, $X$ is an $m\times m$ matrix where we denote by 
$m = \prod_{j = 1}^nr_j$, the size of $D$. Note that both the cardinality of
$L$ and ${\cal I}$ are $m$, and $X$ is non-singular (Theorem 26 of
\cite{Pistone-Riccomagno-Wynn-2001}). 

From these materials, 
the indicator function can be constructed as follows. Let $F =
\{\Bd_{\Bi} \in D\ :\ \Bi \in {\cal I}'\}$
be a fractional factorial design for a subset ${\cal I}' \subset {\cal
I}$. Then the indicator function of $F$ is written uniquely as
\begin{equation}
 f(x_1,\ldots,x_n) = \sum_{\Ba \in L}\theta_{\Ba}\Bx^{\Ba}, 
\label{eqn:indicator-function-form}
\end{equation}
where an $m \times 1$ column vector $\Btheta = (\theta_{\Ba})_{\Ba \in
L}$ is give by $\Btheta = X^{-1}\By$ for an $m\times 1$ column vector
$\By = (y_{\Bi})_{\Bi \in {\cal I}}$ given by
\begin{equation}
 y_{\Bi} = \left\{
\begin{array}{ll}
1, & \mbox{if}\ \Bi \in {\cal I}',\\
0, & \mbox{if}\ \Bi \in {\cal I}\setminus {\cal I}'.\\
\end{array}\right.
\label{eqn:obs-F}
\end{equation}
In other words, the indicator function of $F$ is the interpolatory
polynomial function on $D$ for the observation $\By$ given in
(\ref{eqn:obs-F}). 

The coefficients of the indicator function satisfy the following 
system of the algebraic equations. 
For a polynomial $f$ of the form 
(\ref{eqn:indicator-function-form}), calculate
\begin{equation}
\sum_{\Ba_1 \in L}\sum_{\Ba_2 \in
L}\theta_{\Ba_1}\theta_{\Ba_2}\Bx^{\Ba_1 + \Ba_2}
\label{eqn:indicator-square} 
\end{equation}
and divide it by $G$,
the reduced Gr\"obner basis of $I(D)$, a unique remainder is obtained in
the form $r = \sum_{\Ba \in L}\mu_{\Ba}\Bx^{\Ba}$. This $r$ is called
the standard form of (\ref{eqn:indicator-square}) with respect to $G$.
Then $f$ is an indicator function of some fractional factorial design if
and only if 
$\{\theta_{\Ba}\ :\ \Ba \in L\}$ satisfies the system of algebraic
equations
\begin{equation}
\theta_{\Ba} = \mu_{\Ba},\ \Ba \in L.
\label{eqn:system-equations}
\end{equation}
See Proposition 3.1 of \cite{Aoki-2019} for detail.

In addition to the system of the algebraic equations
(\ref{eqn:system-equations}), 
the orthogonality of the designs can be characterized as the constraints
for the coefficients of the corresponding indicator functions as follows.
Following to Chapter 7 of \cite{Wu-Hamada-2009}, 
we call a design $F \subset D$  is orthogonal
of strength $t\ (t \leq n)$, if for any $t$ factors, all possible
combinations of levels appear equally often in $F$. 
To characterize the 
constraints of a given strength for 
$\{\theta_{\Ba\ :\ \Ba \in L}\}$, we introduce a contrast matrix as
follows (Definition 3.4 of \cite{Aoki-2019}). 
We define the $m\times m$ contrast matrix $C$ by
$C^T = \left[\Bone_m\ |\ C_1^T\ |\ C_2^T\ |\ \cdots\ |\ C_n^T\right]$,
where $\Bone_m = (1,\cdots, 1)^T$ is an $m\times 1$ column vector of the
elements $1$'s, and $C_k$ is a $v_k\times m$ matrix where
\[
 v_k = \sum_{J \subset [n], \#J = k}\left(\prod_{j \in J}(r_j - 1)\right).
\]
The set of $m\times 1$ column vector of $C_k^T$ is
\[
 \left\{\Bc_{J(\tilde{\Bi})} = \{c_{J(\tilde{\Bi})}(\Bi)\}_{\Bi \in
 {\cal I}}\ :\ J \subset [n],\ \#J = k,\ \tilde{\Bi} \in \prod_{j \in
 J}[r_j - 1]\right\},
\]
where 
\[
       c_{J(\tilde{i})}(\Bi) = \left\{
\begin{array}{rl}
1, & \Bi_{J} = 1,\\
-1, & \Bi_{J} = \tilde{i} + 1,\\
0, & \mbox{otherwise}
\end{array}
\right.
      \]
for $\#J = 1$, and
\[
       c_{J(\tilde{\Bi})}(\Bi) = \left\{
\begin{array}{rl}
1, & \Bi_{J} = (\tilde{i}_1,\ldots,\tilde{i}_{k-1}, 1),\\
-1, & \Bi_{J} = (\tilde{i}_1,\ldots,\tilde{i}_{k-1}, \tilde{i}_k + 1),\\
0, & \mbox{otherwise}
\end{array}
\right.
      \]
for $\#J \geq 2$. Note that the contrast matrix $C$ is a non-singular
$m\times m$ matrix (Corollary 3.8 of \cite{Aoki-2019}). Then a
polynomial $f$ of the form (\ref{eqn:indicator-function-form}) is an
indicator function of size $s$, orthogonal 
fractional factorial design of strength $t$,
 if and only if $\{\theta_{\Ba}\ :\ \Ba \in L\}$ satisfies
\begin{equation}
\Bone_m^T X\Btheta = s,\ \ C_{\ell}X\Btheta = \Bzero_{v_{\ell}},\ \ \ell
 = 1,\ldots,t,
\label{eqn:constraint-orthogonal}
\end{equation}
where $\Bzero_{v_{\ell}} = (0,\ldots,0)^T$ is a $v_{\ell}\times 1$
column vector of the element $0$'s, 
in addition to (\ref{eqn:system-equations}). See Theorem 3.6 of
\cite{Aoki-2019} for detail. 

Now we can describe the problem that we consider in this paper. Let
$\mathbb{Q}[\Btheta]$ be the polynomial ring for the variables
$\{\theta_{\Ba}\ :\ \Ba \in L\}$. For given size $s$ and strength of
orthogonality $t$, we can define a polynomial ideal of
$\mathbb{Q}[\Btheta]$ from 
(\ref{eqn:system-equations}) and (\ref{eqn:constraint-orthogonal}) by
\begin{equation}
 I = \left<\{\theta_{\Ba} - \mu_{\Ba},\ \Ba \in L\},\ \Bone_m^TX\Btheta - s,\
 \{C_{\ell}X\Btheta,\ \ell = 1,\ldots,t\}\right>.
\label{eqn:ideal-for-theta}
\end{equation}
Then the variety defined by the polynomial ideal $I$, 
\begin{equation}
 V(I) = \{(a_1,\ldots,a_m) \in \mathbb{Q}^m\ |\ f(a_1,\ldots,a_m) = 0,\
 \forall f \in I\}, 
\label{eqn:variety-for-theta}
\end{equation}
corresponds to the set of the coefficients of the indicator functions of
all orthogonal fractional factorial designs with size $s$ and strength
$t$. By definition, the cardinality of $V(I)$ is finite, i.e., $I$ is a 
$0$-dimensional ideal. In addition, all the zeros of $f \in I$ is
rational, i.e., for all $f \in I$, 
\[
f(a_1,\ldots,a_m) = 0,\
(a_1,\ldots,a_m) \in \mathbb{R}^m\ \Rightarrow\ (a_1,\ldots,a_m) \in
\mathbb{Q}^m 
\]
holds, where $\mathbb{R}$ be the field of real numbers.

\section{Prime decomposition of the radicals}
\label{sec:primary-decomposition}
To enumerate points of the variety (\ref{eqn:variety-for-theta}), 
we use the theory of the primary decomposition. We summarize
fundamental theory of the prime decomposition of the radicals in this section.

Let $I \subset \mathbb{Q}[\Btheta]$ be a 
$0$-dimensional
polynomial ideal $I
\subset \mathbb{Q}[\Btheta]$ given in (\ref{eqn:ideal-for-theta}). 
Let $\sqrt{I} = \{f \in \mathbb{Q}[\Btheta] :\ \exists k \in
\mathbb{Z}_{>0},\ f^k \in I\}$ be the radical ideal of $I$. Then $V(I) =
V(\sqrt{I})$ holds. An ideal $P \subset \mathbb{Q}[\Btheta]$ is called a
prime ideal if
\[
 f,g \in \mathbb{Q}[\Btheta],\ fg \in P\ \Rightarrow\ f \in P\ \mbox{or}\
 g \in P
\]
holds. For prime ideals $P_1,\ldots,P_u$ satisfying
\begin{equation}
\sqrt{I} = P_1\cap\cdots\cap P_u
\label{eqn:prime-decomp}
\end{equation}
and $P_i$ is not included in any $P_j\ (j \neq i)$ for $i = 1,\ldots,u$,
the decomposition (\ref{eqn:prime-decomp}) is called a prime
decomposition of $\sqrt{I}$. See Chapter 4 of \cite{Cox-Little-OShea-1992} for
detail on the primary decomposition of ideals. 
If we have the prime
decomposition (\ref{eqn:prime-decomp}) of $\sqrt{I}$, we can obtain the
expression for the variety $V(I)$
as
\[
 V(I) = V(\sqrt{I}) = V(P_1)\cup\cdots\cup V(P_u).
\]
Here we use 
a relation $V(P_i \cap P_j) = V(P_i) \cup V(P_j)$. 

Our approach is to derive the prime decomposition of $\sqrt{I}$. However, a
direct computation seems to be infeasible for our problem shown in
Section \ref{sec:classification}. Therefore we divide the problem as
follows. For our $0$-dimensional ideal $I \subset \mathbb{Q}[\Btheta]$,
suppose 
\begin{equation}
I = I_1 + I_2 
\label{eqn:I=I1+I2}
\end{equation}
holds, where we define
$I_1 + I_2 = \{f_1 + f_2\ :\ f_1 \in I_1,\ f_2 \in I_2\}$.
Then we have a relation $V(I) = V(I_1) \cap V(I_2)$. Suppose we obtain
the prime decomposition of $\sqrt{I_1}$ as $\sqrt{I_1} = \bigcap_k P_k$. In
this case, from the relation $V(I_1) = \bigcup_k V(P_k)$, we have
\[
 V(I) = \left(\bigcup_k V(P_k)\right) \cap V(I_2) = \bigcup_k\left(V(P_k)
 \cap V(I_2)\right) = \bigcup_k V(P_k + I_2).  
\]
Therefore, if we obtain the prime decomposition $\sqrt{P_k + I_2} =
\bigcap_j Q_{kj}$ for each $k$, we have
$V(P_k) \cap
V(I_2) = \bigcup_j V(Q_{kj})$ for each $k$, and we derive the
decomposition
\[
 V(I) = \bigcup_k\bigcup_j V(Q_{kj}).
\]
We see that our problem is solved in this way in Section
\ref{sec:classification}.

\section{Classifications of half orthogonal fractions of $2^4\times 3$
 designs with strength $2$}
\label{sec:classification}
Now we show the computational results for our problem. We consider the
design of $5$ factors with level sets $A_1 = \cdots = A_4 = \{-1,1\}$
and $A_5 = \{-1,0,1\}$. Our aim is to enumerate and classify all the
half orthogonal fractional factorial designs with strength $2$.
Suppose $I \subset \mathbb{Q}[\Btheta]$ be the corresponding polynomial
ideal defined in (\ref{eqn:ideal-for-theta}), where $m = 48,\ s = 24,\ t = 2$. 
All the
computations are done by Risa/Asir (\cite{Risa/Asir}) installed in 
MacBook Pro, 2.3 GHz, Quad-Core, Intel Core i7.
Our computation is described in the following $3$ steps.

\paragraph*{Step 1:\ Preprocessing}\ Using the constraints
(\ref{eqn:constraint-orthogonal}), some of the variables in $\Btheta =
\{\theta_{\Ba}\ :\ \Ba \in L\}$ can be eliminated as follows.
Following to the expression (\ref{eqn:variety-for-theta}), 
we write $\{\theta_{\Ba}\ :\ \Ba \in L\} = (a_1,\ldots,a_m)$ for
simplicity, and 
suppose for some variable, say $a_1$, $a_1 -
g(a_2,\ldots,a_m) \in I$ holds, 
Then we can eliminate $a_1$ and have $V(I) =
\{(g(a_2,\ldots,a_m),a_2,\ldots,a_m)\ :\ (a_2,\ldots,a_m) \in V'\}$,
where $V' = \{f(g(a_2,\ldots,a_m),a_2,\ldots,a_m)\ :\ f(a_1,\ldots,a_m)
\in I\}$. In this way, from $m = 48$ variables, we eliminate $21$
variables and have the variety $V(I) \subset \mathbb{Q}^{27}$.

\paragraph*{Step 2:\ Primary decomposition}
After Step 1, we find that the generating set of the ideal $I$ is
composed of $48$
polynomials. Let this generating set be $G$ and write $I =
\left<G\right>$. 
We divide $G$ to $G = G_1\cup G_2$, where
$G_1$ is the
polynomials with less than or equal to $4$ terms, and $G_2$ is the
polynomials with greater than $4$ terms. 
From these sets, we define $I_1 = \left<G_1\right>$ and $I_2 =
\left<G_2\right>$ in
(\ref{eqn:I=I1+I2}). We find that there are $15$ polynomials in $G_1$, and
$33$ polynomials in $G_2$, respectively. Following to the strategy given 
in Section \ref{sec:primary-decomposition}, first we compute the prime
decomposition of $\sqrt{I_1}$, which 
we obtain the union of the varieties generated by $111$ prime ideals.  
For each prime ideal, we combine $I_2$ and
calculate prime decomposition again. After these computations, we 
obtain the result that the variety $V(I)$ is composed of $35200$ points
by computation of about $2$ hours. 

\paragraph*{Step 3:\ Classification to equivalence classes}
We classify the obtained $35200$ points to equivalence classes. The
group we consider is a group of permutations of levels for each factor
and permutations of factors among $2$-level factors. Let $S \subset
S_{{\cal I}}$ be the group we consider, where $S_{{\cal I}}$ is a group
of permutations of ${\cal I}$. Then the equivalence classes for
$\Btheta$ with respect to $S$ is $[\Btheta] = \{X^{-1}P_gX\Btheta\ :\ g
\in S\}$, where $P_g$ be an $m\times m$ permutation matrix for $g \in
S$. See Proposition 3.11 of \cite{Aoki-2019} for detail. After
classification, we find that the 
$35200$ points are classified into $63$ equivalence classes. 

\ \\

We summarize the obtained $63$ equivalence classes by their orthogonality of
strength $3$. As for the orthogonality of $3$ two-level factors
$x_1,\cdots,x_4$, we calculate $J$-statistics (\cite{Tang-2001}) given by 
\[
 J_s(F) = \sum_{\Bi \in {\cal I}'}\prod_{j \in s}d_{\Bi j}
\]
for $s \subset \{1,2,3,4\}$, where we define $F =
\{\Bd_{\Bi}\ :\ \Bi \in {\cal I}'\}$ be the fractional factorial design. 
Because $d_{\Bi j} \in \{-1,1\}$ for $j \in
\{1,2,3,4\}$ and the orthogonality of strength $2$, $J_{j_1j_2j_3}(F) = 0$
holds if and only if $8$ possible combinations of levels of $x_{j_1}, x_{j_2},
x_{j_3}$ appear equally often (i.e., $3$ times, respectively) in the
design.
Let $J$ be the set 
$J = \{|J_{123}(F)|, |J_{124}(F)|, |J_{134}(F)|,|J_{234}(F)|\}$,  
which is an invariant for the equivalence classes. We
calculate the set $J$, in addition to the number of three pair of
$\{x_1,x_2,x_3,x_4\}$ which is not orthogonal.
As for the orthogonality among  $2$ two-level factors and three-level
factor $x_5$, we simply count the number of three pair of
$\{x_i,x_j,x_5\},\ i,j \in \{1,2,3,4\}$, which is not orthogonal.
We summarize the number of equivalence classes in Table \ref{tbl:summary}.
\begin{table}[htbp]
\begin{center}
\caption{The classification of equivalence classes of the half
 orthogonal fractions of $2^4\times 3$
 designs with strength $2$ \ ($i,j,k \in \{1,2,3,4\}$)}
\label{tbl:summary}
\begin{tabular}{|c|c|c|c|c|c|c|c|}\hline
\multicolumn{2}{|c|}{} & \multicolumn{6}{|l|}{Num of non-orthogonal}\\
\multicolumn{2}{|c|}{} & \multicolumn{6}{|l|}{triplet of $(x_i,x_j,x_5)$}\\ \cline{1-8}
\multicolumn{1}{|c|}{Num of non-orthogonal} &  & & & & & & \\ 
\multicolumn{1}{|l|}{triplet of $(x_i,x_j,x_k)$} & $J$ &\ $0$\ &\ $1$\ &\ $2$\ &
\ $3$\ &\ $4$\ &\ $5$\ \\ \hline
$0$ & $\{0,0,0,0\}$ & $3^a$ & $4$ & $2$ & $3$ & $1$ & $1$ \\ \hline
$1$ & $\{24,0,0,0\}$ & $1^b$ & & & & & \\ \cline{2-8}
    & $\{16,0,0,0\}$ & $1$ & $1$ & & & & \\ \cline{2-8}
    & $\{8,0,0,0\}$  & $4$ & $8$ & $8$ & $5$ & $2$& \\ \cline{1-8}
$2$ & $\{16,8,0,0\}$ & $1$ & $1$ & $1$ & & & \\ \cline{2-8}
    & $\{8,8,0,0\}$  & $1$ & $4$ & $3$ & $1$& & \\ \cline{1-8}
$3$ & $\{8,8,8,0\}$  & $1$ & $1$ & $3$ & $2$& & \\ \cline{1-8}
\end{tabular}
\end{center}
${}^a$\ :\ Orthogonal fractional factorial designs with strength $3$ shown in \cite{Aoki-2019}.\\
${}^b$\ :\ Regular fractional factorial design with the defining
 relation $x_ix_jx_k = \pm 1$.
\end{table}

We give a list of the indicator functions of the
representative elements for the equivalence classes.
In the list below, we use an index ``Type $T_1$(,$J$)-$T_2$'', meaning that
$T_1$ is a number of non-orthogonal triplet of $(x_i,x_j,x_k)$ and 
$T_2$ is a number of non-orthogonal triplet of $(x_i,x_j,x_5)$.

\begin{itemize}
\item Type $0$-$0$
\begin{itemize}
\item[$\circ$] $2$ relations:\ $\frac{1}{2} + \frac{1}{2}x_1x_2x_3x_4$
\item[$\circ$] $6$ relations:\  $\frac{1}{2} - \frac{1}{2}x_1x_2x_3x_4 + x_1x_2x_4x_5^2$
\item[$\circ$] $48$ relations:\ $\frac{1}{2} - \frac{1}{2}x_1x_2x_4 +
     \frac{1}{4}x_1x_2x_4x_5 + \frac{1}{4}x_1x_2x_3x_4x_5 +
      \frac{3}{4}x_1x_2x_4x_5^2 - \frac{1}{4}x_1x_2x_3x_4x_5^2$
\end{itemize}
\item Type $0$-$1$
\begin{itemize}
\item[$\circ$] $72$ relations:\ $\frac{1}{2} + \frac{1}{2}(x_1x_2x_3x_4
	     + x_3x_4x_5 - x_1x_2x_3x_4x_5^2)$
\item[$\circ$] $144$ relations:\ $\frac{1}{2} - \frac{1}{2}x_1x_2 +
      \frac{3}{4}x_1x_2x_5^2 + \frac{1}{4}(x_1x_2x_3x_5 - x_1x_2x_4x_5 +
      x_1x_2x_3x_4x_5^2)$
\item[$\circ$] $288$ relations:\ $\frac{1}{2} - \frac{1}{2}x_1x_2x_4
+ \frac{1}{4}(x_2x_4x_5 + x_2x_3x_4x_5 - x_1x_2x_3x_4x_5^2) +
	     \frac{3}{4}x_1x_2x_4x_5^2$
\item[$\circ$] $288$ relations:\ $\frac{1}{2} - \frac{1}{2}(x_1x_2x_3x_4 -
      x_1x_2x_3x_4x_5^2) -
      \frac{1}{4}(x_3x_4x_5 - x_1x_3x_4x_5 + x_2x_3x_4x_5 + x_1x_2x_3x_4x_5)$
\end{itemize}
\item Type $0$-$2$
\begin{itemize}
\item[$\circ$] $576$ relations:\ $\frac{1}{2} + \frac{1}{2}(x_1x_2x_3x_4 -
      x_1x_2x_3x_4x_5^2) -
      \frac{1}{4}(x_2x_3x_5 - x_3x_4x_5 + x_1x_2x_3x_5 + x_1x_3x_4x_5)$
\item[$\circ$] $1152$ relations:\ $\frac{1}{2} + \frac{1}{4}(x_1x_2 - x_1x_3 +
      x_1x_2x_4 + x_1x_3x_4 - x_1x_2x_3x_5 - x_1x_2x_3x_4x_5^2) +
	     \frac{1}{8}(x_1x_2x_5  +
      x_1x_3x_5 + x_1x_2x_4x_5 - x_1x_3x_4x_5)
      - \frac{3}{8}(x_1x_2x_5^2 - x_1x_3x_5^2 + x_1x_2x_4x_5^2 +
	     x_1x_3x_4x_5^2)$
\end{itemize}
\item Type $0$-$3$
\begin{itemize}
\item[$\circ$] $192$ relations:\ $\frac{1}{2} -
	     \frac{1}{2}(x_1x_2x_3x_4 - x_1x_2x_3x_4x_5^2) -
      \frac{1}{4}(x_1x_4x_5 - x_2x_4x_5 + x_3x_4x_5 + x_1x_2x_3x_4x_5)$
\item[$\circ$] $288$ relations:\ $\frac{1}{2} + \frac{1}{2}x_1x_2 -
      \frac{3}{4}x_1x_2x_5^2 + \frac{1}{4}(x_2x_3x_5 - x_2x_4x_5 -
	     x_1x_2x_3x_4x_5^2)$
\item[$\circ$] $1152$ relations:\ $\frac{1}{2} + \frac{1}{4}(x_1x_2 - x_1x_3 +
      x_1x_2x_4 + x_1x_3x_4 + x_1x_4x_5 - x_1x_2x_3x_4x_5^2) +
	     \frac{1}{8}(x_1x_2x_5 -
      x_1x_3x_5 - x_1x_2x_4x_5 - x_1x_3x_4x_5)
      - \frac{3}{8}(x_1x_2x_5^2 - x_1x_3x_5^2 + x_1x_2x_4x_5^2 +
	     x_1x_3x_4x_5^2)$ 
\end{itemize}
\item Type $0$-$4$
\begin{itemize}
\item[$\circ$] $144$relations:\ $\frac{1}{2} - \frac{1}{2}(x_1x_2x_3x_4
	     - x_1x_2x_3x_4x_5^2) - \frac{1}{4}(
      x_1x_2x_5 - x_1x_4x_5 + x_2x_3x_5 + x_3x_4x_5)$
\end{itemize}
\item Type $0$-$5$
\begin{itemize}
\item[$\circ$] $576$ relations:\ $\frac{1}{2} -  \frac{1}{4}(x_1x_2 -
	     x_1x_3 + x_1x_4  +
      x_1x_2x_3x_4 - x_2x_3x_5 + x_3x_4x_5) - \frac{1}{8}(x_1x_2x_5  +
      x_1x_3x_5 + x_1x_4x_5 + x_1x_2x_3x_4x_5 - x_1x_2x_3x_4x_5^2)
      + \frac{3}{8}(x_1x_2x_5^2 - x_1x_3x_5^2 + x_1x_4x_5^2)$
\end{itemize}
\item Type $1,\{24,0,0,0\}$-$0$
\begin{itemize}
\item[$\circ$] $8$relations:\ $\frac{1}{2} + \frac{1}{2}x_1x_2x_4$
\end{itemize}
\item Type $1,\{16,0,0,0\}$-$0$
\begin{itemize}
\item[$\circ$] $48$ relations:\ $\frac{1}{2} - \frac{1}{2}x_1x_2x_4 -
      \frac{1}{4}(x_1x_2x_4x_5 - x_1x_2x_3x_4x_5 - x_1x_2x_4x_5^2 +
	     x_1x_2x_3x_4x_5^2)$
\end{itemize}
\item Type $1,\{16,0,0,0\}$-$1$
\begin{itemize}
\item[$\circ$] $288$ relations:\ $\frac{1}{2} - \frac{1}{2}x_1x_2x_4
- \frac{1}{4}(x_2x_4x_5 - x_2x_3x_4x_5 - x_1x_2x_4x_5^2 +
	     x_1x_2x_3x_4x_5^2)$
\end{itemize}
\item Type $1,\{8,0,0,0\}$-$0$
\begin{itemize}
\item[$\circ$] $24$ relations:\ $\frac{1}{2} - \frac{1}{2}x_1x_2x_4 +
	     x_1x_2x_4x_5^2$
\item[$\circ$] $48$ relations:\ $\frac{1}{2} - \frac{1}{2}(x_1x_2x_4 -
	     x_1x_2x_4x_5^2 + x_1x_2x_3x_4x_5^2)$
\item[$\circ$] $48$ relations:\ $\frac{1}{2} - \frac{1}{2}(x_1x_2x_4 -
	     x_1x_2x_3x_4x_5 - x_1x_2x_4x_5^2)$
\item[$\circ$] $144$ relations:\ $\frac{1}{2} - \frac{1}{2}(x_1x_2x_4 -
	     x_2x_3x_4x_5 - x_1x_2x_4x_5^2)$
\end{itemize}
\item Type $1,\{8,0,0,0\}$-$1$
\begin{itemize}
\item[$\circ$] $144$ relations:\ $\frac{1}{2} - \frac{1}{2}(x_1x_2x_4
	     - x_3x_4x_5 - x_1x_2x_4x_5^2)$
\item[$\circ$] $144$ relations:\ $\frac{1}{2} + \frac{1}{2}(x_1x_2x_4 +
	     x_2x_4x_5- x_1x_2x_4x_5^2)$
\item[$\circ$] $288$ relations:\ $\frac{1}{2} - \frac{1}{2}x_1x_2x_4 +
      \frac{1}{4}(x_1x_4x_5 + x_1x_2x_3x_4x_5 - x_1x_3x_4x_5^2) +
	     \frac{3}{4}x_1x_2x_4x_5^2$
\item[$\circ$] $288$ relations:\ $\frac{1}{2} + \frac{1}{2}x_1x_2x_3x_4 -
      \frac{1}{4}(x_3x_4x_5 - x_2x_3x_4x_5 + x_1x_3x_4x_5^2) -
	     \frac{3}{4}x_1x_2x_3x_4x_5^2$
\item[$\circ$] $288$ relations:\ $\frac{1}{2} + \frac{1}{2}x_1x_2x_3x_4 +
      \frac{1}{4}(x_3x_4x_5 + x_2x_3x_4x_5 - x_1x_3x_4x_5^2 -
	     x_1x_2x_3x_4x_5^2)$
\item[$\circ$] $288$ relations:\ $\frac{1}{2} - \frac{1}{2}x_1x_2 +
      \frac{3}{4}x_1x_2x_5^2 + \frac{1}{4}(x_1x_2x_4x_5 +
      x_1x_2x_3x_4x_5 - x_1x_2x_3x_5^2)$
\item[$\circ$] $576$ relations:\ $\frac{1}{2} + \frac{1}{2}(x_1x_2x_4 -
	     x_1x_2x_4x_5^2)
- \frac{1}{4}(x_3x_4x_5 - x_1x_3x_4x_5 + x_2x_3x_4x_5 + x_1x_2x_3x_4x_5)$
\item[$\circ$] $576$ relations:\ $\frac{1}{2} - \frac{1}{2}(x_1x_2x_4 -
	     x_1x_2x_4x_5^2)
+ \frac{1}{4}(x_1x_4x_5 + x_1x_2x_4x_5 - x_1x_3x_4x_5 + x_1x_2x_3x_4x_5)$
\end{itemize}
\item Type $1,\{8,0,0,0\}$-$2$
\begin{itemize}
\item[$\circ$] $576$ relations:\ $\frac{1}{2} - \frac{1}{2}x_1x_2x_4
+ \frac{1}{4}(x_2x_4x_5 + x_3x_4x_5 - x_1x_3x_4x_5^2) +
	     \frac{3}{4}x_1x_2x_4x_5^2$
\item[$\circ$] $576$ relations:\ $\frac{1}{2} - \frac{1}{2}(x_1x_2x_4 -
	     x_1x_2x_4x_5^2)
- \frac{1}{4}(x_2x_3x_5 - x_3x_4x_5 + x_1x_2x_3x_5
      + x_1x_3x_4x_5)$
\item[$\circ$] $576$ relations:\ $\frac{1}{2} + \frac{1}{2}(x_1x_2x_4  -
	     x_1x_2x_4x_5^2)+
      \frac{1}{4}(x_1x_2x_5 + x_2x_4x_5 + x_1x_2x_3x_5 - x_2x_3x_4x_5)$
\item[$\circ$] $576$ relations:\ $\frac{1}{2} - \frac{1}{2}x_1x_2 +
      \frac{3}{4}x_1x_2x_5^2 + \frac{1}{4}(x_2x_4x_5 + x_1x_2x_3x_5 -
	     x_2x_3x_4x_5^2)$
\item[$\circ$] $576$ relations:\ $\frac{1}{2} - \frac{1}{2}x_1x_2 +
      \frac{3}{4}x_1x_2x_5^2 + \frac{1}{4}(x_2x_4x_5 - x_2x_3x_4x_5 +
	     x_1x_2x_3x_5^2)$
\item[$\circ$] $1152$ relations:\ $\frac{1}{2} - \frac{1}{2}(x_1x_2x_4
	     - x_1x_2x_4x_5^2)
+ \frac{1}{4}(x_2x_4x_5 + x_3x_4x_5 + x_1x_2x_4x_5
      - x_1x_3x_4x_5)$
\item[$\circ$] $1152$ relations:\ $\frac{1}{2} + \frac{1}{4}(x_1x_2 - x_1x_3 +
      x_1x_2x_4 + x_1x_3x_4 + x_1x_2x_3x_4x_5 - x_1x_2x_3x_5^2) -
	     \frac{1}{8}(x_1x_2x_5 + x_1x_3x_5 +
      x_1x_2x_4x_5 - x_1x_3x_4x_5)
      - \frac{3}{8}(x_1x_2x_5^2 - x_1x_3x_5^2 + x_1x_2x_4x_5^2 +
	     x_1x_3x_4x_5^2)$
\item[$\circ$] $2304$ relations:\ $\frac{1}{2} + \frac{1}{4}(x_1x_2 - x_1x_3 +
      x_1x_2x_4 + x_1x_3x_4 + x_1x_2x_3x_5 + x_1x_2x_3x_4x_5) +
	     \frac{1}{8}(x_1x_2x_5  +
      x_1x_3x_5 - x_1x_2x_4x_5 - x_1x_3x_4x_5 - x_1x_2x_4x_5^2)
      - \frac{3}{8}(x_1x_2x_5^2 - x_1x_3x_5^2 + x_1x_3x_4x_5^2)$
\end{itemize}
\item Type $1,\{8,0,0,0\}$-$3$
\begin{itemize}
\item[$\circ$] $192$ relations:\ $\frac{1}{2} - \frac{1}{2}(x_1x_2x_4 -
	     x_1x_2x_4x_5^2) - \frac{1}{4}(x_1x_3x_5 - x_2x_3x_5 -
	     x_3x_4x_5 - x_1x_2x_3x_4x_5)$
\item[$\circ$] $576$ relations:\ $\frac{1}{2} + \frac{1}{2}(x_1x_2x_4 -
	     x_1x_2x_4x_5^2)
- \frac{1}{4}(x_1x_4x_5 - x_2x_4x_5 + x_3x_4x_5 + x_1x_2x_3x_4x_5)$
\item[$\circ$] $1152$ relations:\ $\frac{1}{2} - \frac{1}{4}(x_1x_2 - x_1x_3 +
      x_1x_2x_4 + x_1x_3x_4 + x_2x_3x_5 + x_2x_3x_4x_5^2) -
	     \frac{1}{8}(x_1x_2x_5 + x_1x_3x_5 + x_1x_2x_4x_5 - x_1x_3x_4x_5) 
      + \frac{3}{8}(x_1x_2x_5^2 - x_1x_3x_5^2 + x_1x_2x_4x_5^2 +
	     x_1x_3x_4x_5^2)$
\item[$\circ$] $1152$ relations:\ $\frac{1}{2} + \frac{1}{4}(x_1x_2 - x_1x_3 +
      x_1x_2x_4 + x_1x_3x_4 + x_1x_4x_5 - x_1x_2x_3x_5^2) -
	     \frac{1}{8}(x_1x_2x_5 + x_1x_3x_5 - x_1x_2x_4x_5 + x_1x_3x_4x_5)
      - \frac{3}{8}(x_1x_2x_5^2 - x_1x_3x_5^2 + x_1x_2x_4x_5^2 +
	     x_1x_3x_4x_5^2)$
\item[$\circ$] $2304$ relations:\ $\frac{1}{2} + \frac{1}{4}(x_1x_2 - x_1x_3 +
      x_1x_2x_4 + x_1x_3x_4 + x_1x_4x_5 + x_1x_2x_3x_4x_5) +
	     \frac{1}{8}(x_1x_2x_5  -
      x_1x_3x_5 + x_1x_2x_4x_5 - x_1x_3x_4x_5 - x_1x_2x_4x_5^2)
      - \frac{3}{8}(x_1x_2x_5^2 - x_1x_3x_5^2 + x_1x_3x_4x_5^2)$
\end{itemize}
\item Type $1,\{8,0,0,0\}$-$4$
\begin{itemize}
\item[$\circ$] $576$ relations:\ $\frac{1}{2} + \frac{1}{2}(x_1x_2x_4 -
	     x_1x_2x_4x_5^2) - \frac{1}{4}(x_1x_2x_5 - x_1x_4x_5 +
	     x_2x_3x_5 + x_3x_4x_5)$
\item[$\circ$] $1152$ relations:\ $\frac{1}{2} + \frac{1}{4}(x_1x_2 -
	     x_1x_3 +
      x_1x_2x_4 + x_1x_3x_4 + x_2x_4x_5 + x_3x_4x_5) +
      \frac{1}{8}(x_1x_2x_5  - x_1x_3x_5 + x_1x_2x_4x_5 - x_1x_3x_4x_5 -
      x_1x_2x_4x_5^2)
      - \frac{3}{8}(x_1x_2x_5^2 - x_1x_3x_5^2 + x_1x_3x_4x_5^2)$
\end{itemize}
\item Type $2,\{16,8,0,0\}$-$0$
\begin{itemize}
\item[$\circ$] $144$relations:\ $\frac{1}{2} - \frac{1}{2}(x_1x_2x_4 -
	     x_1x_2x_4x_5^2 + x_1x_3x_4x_5^2)$
\end{itemize}
\item Type $2,\{16,8,0,0\}$-$1$
\begin{itemize}
\item[$\circ$] $288$ relations:\ $\frac{1}{2} - \frac{1}{2}x_1x_2x_4 -
      \frac{1}{4}(x_1x_4x_5 - x_1x_2x_3x_4x_5 - x_1x_2x_4x_5^2 +
	     x_1x_3x_4x_5^2)$
\end{itemize}
\item Type $2,\{16,8,0,0\}$-$2$
\begin{itemize}
\item[$\circ$] $576$ relations:\ $\frac{1}{2} - \frac{1}{2}x_1x_2x_4
- \frac{1}{4}(x_2x_4x_5 - x_3x_4x_5 - x_1x_2x_4x_5^2 + x_1x_3x_4x_5^2)$
\end{itemize}
\item Type $2,\{8,8,0,0\}$-$0$
\begin{itemize}
\item[$\circ$] $288$ relations:\ $\frac{1}{2} - \frac{1}{2}(x_1x_2x_4 -
	     x_1x_2x_4x_5^2)
      -  
      \frac{1}{4}(x_1x_3x_4x_5 - x_1x_2x_3x_4x_5 + x_1x_3x_4x_5^2 +
	     x_1x_2x_3x_4x_5^2)$
\end{itemize}
\item Type $2,\{8,8,0,0\}$-$1$
\begin{itemize}
\item[$\circ$] $144$ relations:\ $\frac{1}{2} - \frac{1}{2}x_1x_2 +
      \frac{3}{4}x_1x_2x_5^2 + \frac{1}{4}(x_1x_2x_3x_5^2 -
      x_1x_2x_4x_5^2 + x_1x_2x_3x_4x_5^2)$
\item[$\circ$] $288$ relations:\ $\frac{1}{2} + \frac{1}{2}(x_1x_2x_3x_4 -
      x_1x_2x_3x_4x_5^2) -
      \frac{1}{4}(x_1x_3x_5 - x_1x_2x_3x_4x_5 + x_1x_2x_3x_5^2 +
	     x_1x_3x_4x_5^2)$
\item[$\circ$] $576$ relations:\ $\frac{1}{2} - \frac{1}{2}(x_1x_2x_4 -
	     x_1x_2x_4x_5^2)
- \frac{1}{4}(x_1x_4x_5 - x_1x_2x_4x_5 + x_1x_3x_4x_5^2 + x_1x_2x_3x_4x_5^2)$
\item[$\circ$] $1152$ relations:\ $\frac{1}{2} - \frac{1}{2}(x_1x_2x_4
	     - x_1x_2x_4x_5^2)
- \frac{1}{4}(x_3x_4x_5 - x_2x_3x_4x_5 + x_1x_3x_4x_5^2 + x_1x_2x_3x_4x_5^2)$
\end{itemize}
\item Type $2,\{8,8,0,0\}$-$2$
\begin{itemize}
\item[$\circ$] $576$ relations:\ $\frac{1}{2} + \frac{1}{2}(x_1x_2x_3x_4
	     - x_1x_2x_3x_4x_5^2) +
      \frac{1}{4}(-x_2x_3x_5 + x_3x_4x_5 - x_1x_2x_3x_5^2 - x_1x_3x_4x_5^2)$
\item[$\circ$] $1152$ relations:\ $\frac{1}{2} + \frac{1}{4}(x_1x_2 - x_1x_3 +
      x_1x_2x_4 + x_1x_3x_4 - x_1x_2x_3x_5 - x_1x_2x_3x_4x_5^2) -
	     \frac{1}{8}(x_1x_2x_5 +
      x_1x_3x_5 - x_1x_2x_4x_5 + x_1x_3x_4x_5 + x_1x_2x_4x_5^2 + x_1x_3x_4x_5^2)
      - \frac{3}{8}(x_1x_2x_5^2 - x_1x_3x_5^2)$
\item[$\circ$] $2304$ relations:\ $\frac{1}{2} - \frac{1}{4}(x_1x_2 -
	     x_1x_3 +
      x_1x_2x_4 + x_1x_3x_4 - x_1x_2x_3x_5^2 - x_1x_2x_3x_4x_5^2) +
	     \frac{1}{8}(- x_1x_2x_5 - x_1x_3x_5 +
      x_1x_2x_4x_5 + x_1x_3x_4x_5 + x_1x_2x_4x_5^2)
      + \frac{3}{8}(x_1x_2x_5^2 - x_1x_3x_5^2 + x_1x_3x_4x_5^2)$ 
\end{itemize}
\item Type $2,\{8,8,0,0\}$-$3$
\begin{itemize}
\item[$\circ$] $1152$ relations:\ $\frac{1}{2} + \frac{1}{4}(x_1x_2 - x_1x_3 +
      x_1x_2x_4 + x_1x_3x_4 - x_1x_4x_5 - x_1x_2x_3x_4x_5^2) -
	     \frac{1}{8}(x_1x_2x_5 -
      x_1x_3x_5 + x_1x_2x_4x_5 + x_1x_3x_4x_5 + x_1x_2x_4x_5^2 + x_1x_3x_4x_5^2)
      - \frac{3}{8}(x_1x_2x_5^2 - x_1x_3x_5^2)$
\end{itemize}
\item Type $3$-$0$
\begin{itemize}
\item[$\circ$] $192$ relations:\ $\frac{1}{2} - \frac{1}{2}(x_1x_2x_4 -
	     x_1x_2x_4x_5^2)
+ \frac{1}{4}(x_1x_3x_4x_5 + x_2x_3x_4x_5 - x_1x_3x_4x_5^2 + x_2x_3x_4x_5^2)$
\end{itemize}
\item Type $3$-$1$
\begin{itemize}
\item[$\circ$] $576$ relations:\ $\frac{1}{2} - \frac{1}{2}(x_1x_2x_4 +
	     x_1x_2x_4x_5^2) - 
      \frac{1}{4}(x_1x_3x_5 - x_1x_2x_3x_4x_5 + x_1x_2x_3x_5^2 +
	     x_1x_3x_4x_5^2)$
\end{itemize}
\item Type $3$-$2$
\begin{itemize}
\item[$\circ$] $576$ relations:\ $\frac{1}{2} - \frac{1}{2}(x_1x_2x_4 -
	     x_1x_2x_4x_5^2)
- \frac{1}{4}(x_2x_3x_5 - x_3x_4x_5 + x_1x_2x_3x_5^2 + x_1x_3x_4x_5^2)$
\item[$\circ$] $576$ relations:\ $\frac{1}{2} + \frac{1}{2}(x_1x_2x_4 -
	     x_1x_2x_4x_5^2) +
      \frac{1}{4}(x_1x_2x_5 + x_2x_4x_5 + x_1x_2x_3x_5^2 -
	     x_2x_3x_4x_5^2)$
\item[$\circ$] $1152$ relations:\ $\frac{1}{2} + \frac{1}{4}(x_1x_2 -
	     x_1x_3 + x_1x_2x_4
      + x_1x_3x_4 + x_1x_2x_3x_4x_5 - x_1x_2x_3x_5^2) +
	     \frac{1}{8}(x_1x_2x_5  + x_1x_3x_5 - x_1x_2x_4x_5 +
      x_1x_3x_4x_5 - x_1x_2x_4x_5^2 - x_1x_3x_4x_5^2)
      - \frac{3}{8}(x_1x_2x_5^2 - x_1x_3x_5^2)$
\end{itemize}
\item Type $3$-$3$
\begin{itemize}
\item[$\circ$] $384$ relations:\ $\frac{1}{2} - \frac{1}{4}(x_1x_2 - x_1x_3 +
      x_1x_2x_4 + x_1x_3x_4 + x_2x_3x_5 + x_2x_3x_4x_5^2) +
	     \frac{1}{8}(x_1x_2x_5 +
      x_1x_3x_5 - x_1x_2x_4x_5 + x_1x_3x_4x_5 + x_1x_2x_4x_5^2 + x_1x_3x_4x_5^2)
      + \frac{3}{8}(x_1x_2x_5^2 - x_1x_3x_5^2)$ 
\item[$\circ$] $384$ relations:\ $\frac{1}{2} + \frac{1}{4}(x_1x_2 - x_1x_3 +
      x_1x_2x_4 + x_1x_3x_4 + x_1x_4x_5 - x_1x_2x_3x_5^2) +
	     \frac{1}{8}(x_1x_2x_5  +
      x_1x_3x_5 + x_1x_2x_4x_5 - x_1x_3x_4x_5 - x_1x_2x_4x_5^2 - x_1x_3x_4x_5^2)
      - \frac{3}{8}(x_1x_2x_5^2 - x_1x_3x_5^2)$
\end{itemize}
\end{itemize}

\section{Discussion}
\label{sec:discussion}
In this paper, we consider using the theory of primary decomposition to
enumerate fractional factorial designs with given orthogonality. As
shown in Section 4, we have broken through the limit of the previous
work (\cite{Aoki-2019}) in this paper. 
Our approach is to divide generators of the ideal according to the
number of terms. We first compute the primary decomposition of the ideal
generated by the polynomials with less than or equal to $4$ terms, and
then combine the result to the other polynomials. This heuristic
approach works well for our problem. Of course, it may be difficult to
carry out this heuristic approach for problems of larger sizes. 

The enumeration and classification of orthogonal designs
is one of the fundamental problems in design of experiments. 
For practical contribution, we can apply the
classification to the theory of the optimal design. For the result given
in Section 4, however, we see that the simple regular design
$x_1x_2x_3x_4 = \pm 1$ is the optimal, for the criterion such as $D$-,
$A$-, $E$- optimalities under some statistical models. Therefore, 
the contribution of this paper is restricted to the theoretical one.
For more practical contribution, the problems for the designs where
there does not exist regular fractions should be considered, which is
our future work.

\bibliographystyle{plain}

\end{document}